\documentclass[%
 aip,
 jmp,%
 amsmath,amssymb,
reprint,%
]{revtex4-1}

\usepackage{graphicx}
\usepackage{dcolumn}
\usepackage{bm}

\begin{document}


\title{Indirect-to-direct band gap crossover of single walled MoS$_2$ nanotubes}
\author{Kaoru Hisama}
\affiliation{%
 Department of Mechanical Engineering, The University of Tokyo,\\
 Tokyo 113-8656, Japan
}%
\author{Mina Maruyama}
\affiliation{
 Department of Physics, University of Tsukuba,
 Tsukuba 305-8571, Japan
}%
\author{Shohei Chiashi}
\affiliation{%
 Department of Mechanical Engineering, The University of Tokyo,\\
 Tokyo 113-8656, Japan
}%
\author{Shigeo Maruyama}
 \email{maruyama@photon.t.u-tokyo.ac.jp}
\affiliation{%
 Department of Mechanical Engineering, The University of Tokyo,\\
 Tokyo 113-8656, Japan
}%

\author{Susumu Okada}
 \email{sokada@comas.frsc.tsukuba.ac.jp}
\affiliation{
 Department of Physics, University of Tsukuba,
 Tsukuba 305-8571, Japan
}%

\date{\today}

\begin{abstract}
Using density functional theory, the electronic structures of single walled molybdenum disulfide nanotubes (MoS$_2$ NTs) were investigated as a function of diameter. Our calculations show that the electronic structure near the band gap is sensitive to the NT diameter: armchair MoS$_2$ NTs act as indirect gap semiconductors for diameters up to approximately 5.0 nm, while armchair MoS$_2$ NTs with larger diameters act as direct gap semiconductors with band edges located in the vicinity of $k = 2\pi/3$. This finding implies that MoS$_2$ NTs with large diameters should exhibit similar photoluminescence to 2D monolayer MoS$_2$ sheets. This indirect-to-direct band gap crossover is ascribed to the upward shift of the valence band peak at the $\Gamma$ point in small diameter NTs, which is caused by the tensile strain resulting from their tubular structures.
\end{abstract}

\maketitle

Since the discovery of graphene, atomic layer materials have attracted much attention due to their unique physical properties arising from chemically inert 2D covalent networks with an unusual network topology. The constituent elements and network structure provide these materials with a variety of properties, ranging from metals to insulators. Transition metal dichalcogenides (TMDCs), such as MoS$_2$, WS$_2$, WSe$_2$, and MoTe$_2$, are representative examples of such atomic layer materials. TMDCs consist of an atomic layer of transition metals that form a triangular lattice sandwiched by atomic layers of chalcogens arranged in a prismatic manner, resulting in a hexagonal network of these elements with a thickness of approximately 0.3 nm \cite{Friend1987,Wang2012}. 
According to the chemical valence of the constituent elements in TMDCs, monolayers of TMDCs are primarily semiconductors, with a direct band gap at the K point \cite{Roldan2014,Mak2010, Splendiani2010}, which strongly depends on the constituent elements; in contrast, corresponding thin films or bulk materials are indirect band gap semi-conductors.

Due to the chemical stability of these 2D lattices with covalent bonds, TMDCs have been employed as starting materials to construct various derivatives. Such derivatives can provide building blocks of various layered heterostructures, where each layer is bound via weak van der Waals interactions \cite{Geim2013,Novoselov2016,Liu2016}. In these van der Waals heterostructures, one can tune the electronic structure near the band edges by assembling and stacking them in an appropriate manner, leading to the potential for optical and electronic applications \cite{Lee2014,Zhang2016}. Nanoribbons provide another possible derivative, obtained by imposing an open boundary condition on TMDC sheets. The formation energy and electronic structures of TMDC nanoribbons are strongly dependent on the edge morphology \cite{Li2008, Ataca2011}, due to the polarity at the edges and structural relaxation. In addition to nanoribbons, nanotubes (NTs) of TMDCs \cite{Tenne1992,Tenne2010} are also available, by imposing a tubular boundary condition on nanoribbons. Indeed, researchers recently reported the synthesis of single walled MoS$_2$ NT wrapping walls of multi-walled boron nitride NTs \cite{Xiang2020}. The authors reported photoluminescence (PL) from these single walled MoS$_2$ NTs, with a similar energy to the 2D monolayer MoS$_2$. This behavior implies that the electronic properties of single walled MoS$_2$ NT act as a direct band gap semiconductor, even with the tubular structure.

Prior to the experimental synthesis of single walled MoS$_2$ NTs, their electronic properties were intensively studied \cite{Seifert2000, Zhang2010, Zibouche2012}. It has been reported that these NTs have intrinsic semiconducting electronic structures because their 2D counterpart is a semiconductor with a direct band gap at the K point \cite{Mak2010, Splendiani2010}. It has also been revealed that the band edge structure depends on the NT chirality. 
Zigzag ($n$,0) NTs have direct band gaps at the $\Gamma$ point, while armchair ($n$,$n$) and chiral NTs have indirect band gaps between the vicinity of the $\Gamma$ point and the middle of the 1D Brillouin zone (BZ) for the valence and conduction band edges, respectively. These theoretical insights into the electronic band structure of single-walled MoS$_2$ NTs are not fully accountable for the recent experimental finding that MoS$_2$ NTs have a direct band gap causing PL. In addition, the results also imply the existence of a missing link that connects the indirect band gap of the tubular form and the direct band gap of the planar form of MoS$_2$.

In this report, we aim to determine the critical NT diameter at which the indirect-to-direct band gap crossover occurs in order to connect the discontinuity between the indirect gaps of NTs and the direct gap of sheets. Using density functional theory (DFT) \cite{Hohenberg1964, Kohn1965}, we investigated the electronic structures of armchair MoS$_2$ NTs, with diameters ranging from 1.6 to 6.8 nm. The DFT calculations revealed that MoS$_2$ NTs with diameters larger than 5.2 nm are semiconductors with an indirect band gap in the vicinity of $k = 2\pi/3$, corresponding to the valence and conduction band valleys at the K point of an isolated MoS$_2$ sheet. Furthermore, our structural analysis of MoS$_2$ NTs demonstrated that the curvature-induced mechanical strain causes the indirect band gaps of MoS$_2$ NTs with small diameters.


All of the calculations in this study were performed using DFT implemented in the Simulation Tool for Atom Technology (STATE) program package \cite{Morikawa2001}. The general gradient approximation (GGA) was applied with the Perdew–Burke–Ernzerhof functional \cite{Perdew1996, Perdew1997} to describe the exchange correlation potential energy among interacting electrons. The ultrasoft pseudopotential generated by the Vanderbilt scheme \cite{Vanderbilt1990} was used to provide the interaction between valence electrons and nuclei. The valence wave functions and deficit charge density were expanded in terms of the plane wave basis set, with cutoff energies of 340 and 3061 eV (25 and 225 Ry), respectively.

In this work, we considered armchair MoS$_2$ NTs with diameters ranging from 1.6 to 6.8 nm, which correspond to the indices (9,9) and (39,39), to determine the critical diameter at which the band gap crossover occurs. To exclude the effect arising from inter-tube periodic boundary conditions normal to the NT, the NTs were separated from their periodic images by 0.9 nm vacuum spacing. BZ integration was conducted under six equidistant k points along the tube axis. The atomic coordinates were optimized using a fixed lattice parameter of 0.315 nm, corresponding to the experimental value for planar MoS$_2$ \cite{Wakabayashi1975}, until the forces acting on each atom were less than 0.684 eV/nm (1.33$\times$10$^{-3}$ Hartree/au).

\begin{figure}
\centering
\includegraphics[width=0.45\textwidth]{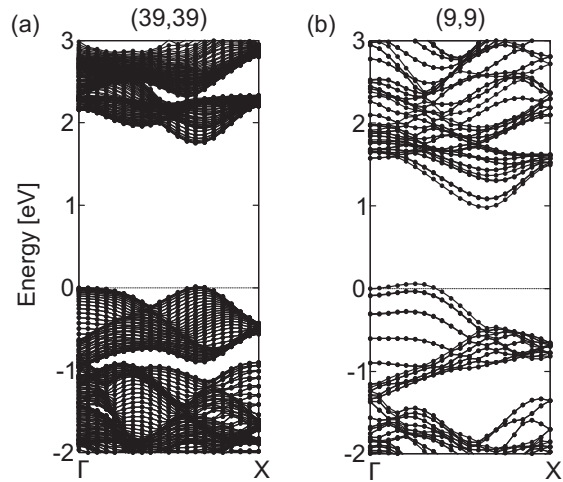}
\caption{Electronic structures of (a) (39,39) MoS$_2$ NT and (b) (9,9) MoS$_2$ NT. The energies are measured with respect to that of the valence band edge at the $\Gamma$ point. }
\label{fig:electronic_structure}
\end{figure}

Figure \ref{fig:electronic_structure} shows the electronic structure of the MoS$_2$ NTs, from the $\Gamma$ point ($k=0$) to the X point ($k=\pi$) in the Brillouin zone. The band dispersion relation was sensitive to the tube diameter, even though these NTs had the same armchair chirality. The (39,39) MoS$_2$ NT was a direct band gap semiconductor, with the valence and conduction band edges located in the vicinity of $k=2\pi/3$ [Fig. \ref{fig:electronic_structure}(a)], corresponding to the K point of the 2D BZ. This result indicates that the band edge positions of the NTs with large diameters originate from the valleys of the isolated MoS$_2$ sheet, confirming the expectation that the electronic band structure near the gap of the NTs should asymptotically approach that of an isolated MoS$_2$ sheet. In contrast to the (39,39) NT, the (9,9) NT had an indirect band gap between near $\Gamma$ and $k=2\pi/3$ for the valence and conduction band edges, respectively [Fig. \ref{fig:electronic_structure}(b)], as observed in the previous calculations. Thus, NTs with a large diameter should exhibit PL, as observed for isolated sheets of MoS$_2$.

\begin{figure}
\centering
\includegraphics[width=0.45\textwidth]{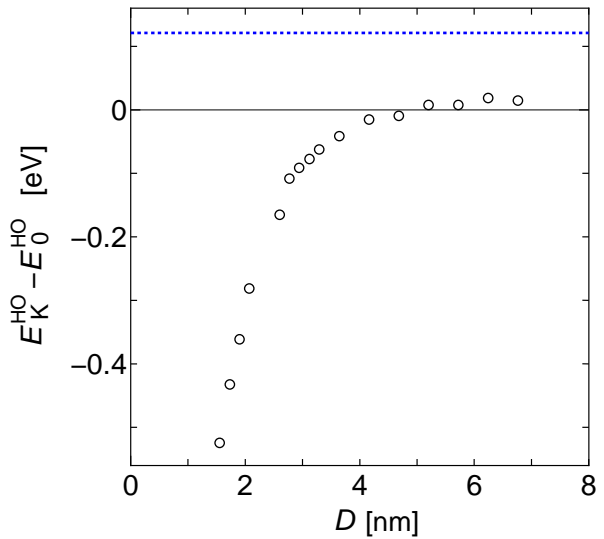}
\caption{Energy difference between valence band peaks, $E_K^{\mathrm{HO}}-E_0^{\mathrm{HO}}$, as a function of diameter, $D$, for MoS$_2$ NTs. The dotted line indicates the value for monolayer MoS$_2$. }
\label{fig:band_gap_difference}
\end{figure}

To determine the critical diameter at which the indirect-to-direct band gap crossover occurs, we investigated the energy difference between the peak of the highest energy of the occupied state in valence band at near $\Gamma$, $E_{0}^{\mathrm{HO}}$, and at $k=2\pi/3$, corresponding to the K point of 2D, $E_{K}^{\mathrm{HO}}$. Figure \ref{fig:band_gap_difference} shows the energy difference, $E_{0}^{\mathrm{HO}}-E_{K}^{\mathrm{HO}}$ as a function of NT diameter, $D$. 
The energy difference, $E_{0}^{\mathrm{HO}}-E_{K}^{\mathrm{HO}}$ monotonically increases with increasing diameter, crossing zero at a diameter of approximately 5.0 nm. Therefore, the NTs with a diameter of 5.0 nm or larger have a direct band gap in the vicinity of $k=2\pi/3$ and are expected to exhibit PL at the band gap energy. Note that the asymptotic property of the energy peak difference rises extremely slowly with increasing diameter. The peak difference for the NT with the largest diameter is still smaller (by 0.1 eV) than that of the monolayer MoS$_2$ sheet.

\begin{figure}
\centering
\includegraphics[width=0.4\textwidth]{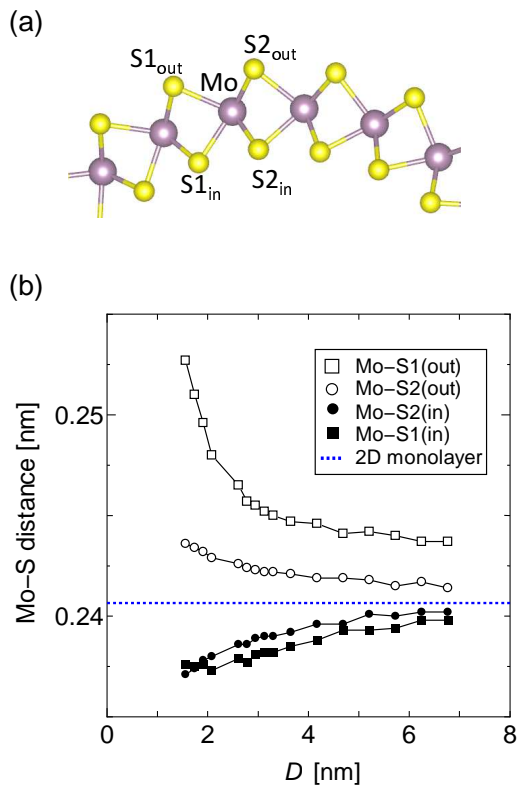}
\caption{
(a) Schematic view of an atomic arrangement of an MoS$_2$ NT, where the yellow and purple balls indicate S and Mo atoms, respectively. (b) Mo-S bond length as a function of the MoS$_2$ NT diameter, $D$, where indices correspond with those in (a).}
\label{fig:atomic_structure}
\end{figure}

The diameter dependence of the valence band edge of the MoS$_2$ NTs indicates that the effect of the structural strain arises from the curvature. Figure \ref{fig:atomic_structure} shows the bond length between Mo and S atoms as a function of tube diameter. The tubular structure and symmetry result in four distinct Mo-S distances. Two of the four distances, corresponding to outer Mo-S bonds, increase with decreasing diameter, while the other two distances, corresponding to inner Mo-S bonds, decrease with decreasing diameter. The bond deformations of the outer Mo-S bonds are larger than those of the inner Mo-S bonds. In particular, the outer Mo-S bond perpendicular to the tube axis is elongated by a factor of four compared with that of the inner Mo-S bonds. Therefore, the atoms in the MoS$_2$ NTs are subjected to tensile and compressive strain in the outer and inner regions, respectively. Furthermore, the tensile strain for the outer part is dominant for the structural distortion. This structural deformation can account for the diameter-dependent edge modulation observed for the valence band of the MoS$_2$ NT because the valence band edge of monolayer MoS$_2$ is sensitive to strain \cite{Conley2013,Gong2013,Ghorbani-Asl2013}.

\begin{figure}
\centering
\includegraphics[width=0.4\textwidth]{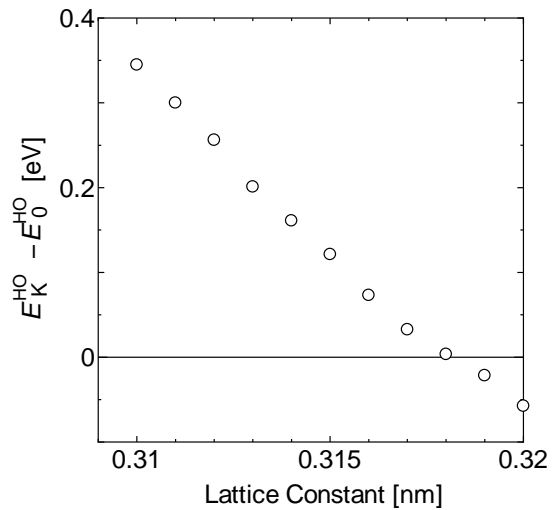}
\caption{ 
Energy difference between valence band peaks, $E_K^{\mathrm{HO}}-E_0^{\mathrm{HO}}$, of monolayer MoS$_2$ as a function of the lattice constant. }
\label{fig:monolayer_strain}
\end{figure}

To confirm the effect of strain on the electronic structure of MoS$_2$ sheets, we investigated the electronic structure of isolated MoS$_2$ sheets. Figure \ref{fig:monolayer_strain} shows the energy difference between the K and $\Gamma$ points of the highest branch of the valence band as a function of the lattice parameter. With increasing lattice parameter, which corresponds to the tensile strain, the energy difference decreased and crossed zero at a tensile strain of 1.84\%, with an Mo-S distance of 0.318 nm. Therefore, the tensile strain caused the upward shift of the peak at the $\Gamma$ point, leading to the indirect band gap structure for the MoS$_2$ sheet. This upward shift is associated with the wave function distribution of the valence band top at the $\Gamma$ point, where the state consists of the $d_z$ state of Mo and the $p_z$ state of S atoms with a bonding nature \cite{Gong2013}. Therefore, the state was shifted upward by the tensile strain, which reduced the orbital hybridization. This analysis clearly confirms that the band gap crossover in MoS$_2$ NTs is induced by the strain associated with their curvature.



Based on DFT with GGA, we investigated the electronic structure of armchair MoS$_2$ NTs with diameters of 1.6-6.8 nm. Our DFT calculations showed that NTs with diameters of up to 5.0 nm are semiconductors with an indirect band gap, whereas NTs with diameters of 5.0 nm or larger have a direct band gap between valleys in the vicinity of $k=2\pi/3$, corresponding to the band edge valleys of monolayer MoS$_2$. The results confirm that the electronic structure of MoS$_2$ exhibits asymptotical properties transitioning from an indirect to a direct band gap based on the diameter. Therefore, MoS$_2$ NTs with diameters of 5.2 nm or larger may exhibit PL, as observed for monolayer MoS$_2$ sheets, and may represent a constituent material for designing optical, optoelectronic, and valleytronic devices. Structural analysis of the MoS$_2$ NTs revealed that the tensile strain on the outer part of the NTs caused an upward shift of the valence band peak near the $\Gamma$ point, leading to the indirect-to-direct band gap crossover associated with the NT diameter.
 

Part of this work was financially supported by JSPS KAKENHI Grant Numbers 	JP17K06187, JP18H05329, JP19J13818 and JP20H00220. 
Some of the calculations were performed on NEC SX-Ace at the Cybermedia Center in Osaka University and on Oakbridge-CX at Information Technology center in the University of Tokyo.

\bibliography{paper5}

\end{document}